# Dispersion in magnetostatic CoTaZr spin wave-guides

A. Kozhanov, D. Ouellette, Z. Griffith, M. Rodwell, and S. J. Allen

California Nanosystems Institute, University of California at Santa Barbara, Santa Barbara, CA, 93106

A. P. Jacob

Technology and Manufacturing Group, Intel Corporation, Santa Clara, CA 95052

D. W. Lee and S. X. Wang

Department of Materials Science and Engineering, Sanford University, Stanford, CA, 94305

Magnetostatic spin wave dispersion and loss are measured in micron scale spin wave-guides in ferromagnetic, metallic CoTaZr. Results are in good agreement with model calculations of spin wave dispersion and up to three different modes are identified. Attenuation lengths of the order of 3 microns are several of orders of magnitude shorter than that predicted from eddy currents in these thin wires.

A current technology drive, directed toward future signal processing and logic devices, attempts to introduce spin degrees of freedom as an alternative, complement or companion to semiconductor charge based electronics. Magnetic bipolar transistors [1], spin metal-oxide-semiconductor field effect transistors (MOSFETs) [2,3] and spin torque transfer devices [4,5,6] explore potential performance enhancement by sensing or using the spin degree of freedom that accompanies the charge current flow [7]. Spin waves can transfer spin information and have the potential for spin control without directly moving charge. For example, spin wave interference could be used in Mach-Zehnder type interferometer as a wave based logical element. [8,9,10] Further, spin waves are intrinsically non-linear[11,12], interaction of the spin wave with a locally controlled magnetization could lead to a spin wave switch.

Microwave devices like delay lines, filters and resonators based on magneto-static waves in insulating ferrimagnetic materials like yttrium-iron garnet (YIG) [13] have long been explored and developed. However, future micro and nano scale spin wave logic devices may benefit from exploiting ferromagnetic metals that are more easily deposited, processed and nanofabricated than ferrimagnetic oxides. Further, ferromagnetic metals like CoTaZr and CoFe have nearly an order of magnitude larger saturation magnetization than typical ferrimagnets.[14] As a result, they will support higher, shape defined, zero magnetic field resonances and consequently intrinsically faster response. Configurations that support spin wave modes without requiring external magnetic fields are essential for future micro/nanodevices. Spin wave damping by eddy currents must be mitigated but dipolar magnetostatic waves are little affected by material discontinuity and micro and nanostructuring can diminish eddy current losses.

This letter describes excitation, detection and propagation of backward volume magnetostatic spin waves (BVMSW) in ferromagnetic wires.[15] The magnetization lies along the axis of the wire due to strong shape anisotropy. In the *absence* of a magnetic field, it has a finite ferromagnetic resonance determined by the cross sectional shape and supports backward volume spin waves propagating along the wire.

Co$_{90}$Zr$_5$Ta$_5$ ferromagnetic films, 110 nm thick, were grown by sputter deposition on a Si/SiO$_2$ substrate. A vibrating sample magnetometer measured a saturation magnetization of $M_S \approx 1.2\,\text{T}$ and a coercive field of $H_C \approx 2\,\text{Oe}$ in the unpatterned film. Wires were produced with ICP dry etch system using chlorine chemistry. An insulating SiO$_2$ layer covered the patterned ferromagnetic wires.

Magnetostatic spin waves were excited and detected by coupling loops formed by the short-circuited ends of coplanar waveguides (CPW) [16] (Fig. 1.). High frequency currents in the signal line of the CPW aligned atop the ferromagnetic wire produced magnetic fields that excite the magnetostatic spin waves as shown in Fig. 2. Because of the high permeability of the CoTaZr perpendicular to the wire, the high frequency magnetic field is largely outside the wire.[17, 18, 19] This causes poor but sufficient coupling to excite and detect the magneto-static waves.

S-parameters were measured at room temperature using an Agilent 8720ES vector network analyzer. Only S$_{12}$, the ratio of high frequency voltage at terminals 1 to the input high frequency voltage at terminals 2 is analyzed in the following discussion. The test devices were positioned on the gap of electro-magnet that provided bias up to 1000 Oe. By comparing the *S*-parameters at disparate bias magnetic fields, the magnetic field independent instrument response can be effectively removed to expose the *S* - parameters of the magneto-static spin wave guide. The central point of this paper is to show that the magnetic field dependent coupling of the two shorted CPW can be described by spin waves.

The real part of the magneto-static spin wave contribution to *S*$_{12}$ is shown in Fig. 3. A strong resonant feature moves to high frequency with increasing magnetic field. At a particular magnetic field, there is very little signal at frequencies below the strong feature whereas at higher frequencies many discernible "noise like" features appear. In the following we successfully interpret these "noise like", but reproducible, features, which occur over a relatively broad frequency range, as coupling of the CPWs by wire spin waves . It should also be noted that the amplitude of the transmitted signal "turns on" with a modest magnetic field and then exhibits little growth with further magnetic bias. Micromagnetic simulations confirm that without a small bias field, ~50 Oe, the magnetization, at the wire ends, under the coupling loops is not well aligned and the coupling to the spin excitations suppressed.

The real and imaginary parts of $S_{12}$ for a (5×1.8×0.11) μm$^3$ CoZrTa wire with a bias of 594 Oe are shown in Fig. 4. Despite the large variations in the amplitude $|S_{12}|$, a polar plot, Fig. 4b, shows the phase making a steady evolution as the frequency increases from 8 to 13 GHz. Despite the "noise like" but *reproducible* amplitude spectrum, the phase appears to wind through ~10 π radians. While the relative phase of $S_{12}$ evolves smoothly when the amplitude is strong, near minima in $S_{12}$ the phase appears to jump by ~ π radians before continuing to wind in a well-defined manner as the transmitted signal grows again as shown in Fig. 4c (we mark these regions with a mode index *p*).

The evolution of the phase of $S_{12}$ suggests that spin waves couple the two CPW. In particular, if we assume that the effective separation between the point of excitation and detection is *r* then the frequency dependent phase, $\varphi_p(f)$, can be simply related to a frequency dependent wave vector, $k(f)$, by

$$\varphi_p(f) \approx k_p(f) r + \varphi_{0p}. \qquad (1)$$

$\varphi_{0p}$ is an unknown phase for any given mode, *p*, and *r* is the length separating excitation and detection. In the analysis that follows we *assume* that $\varphi_{0p}$ is a constant; we ignore any frequency dependent phase change introduced by the coupling of the CPW to the patterned spin wave guide. We take the overall agreement of the experiment and theoretical model as *a posteriori* validation of this assumption.

Following (1), if $\varphi_{0p}$ is a constant, changes in $\varphi_p(f)$ with frequency are due to the change in the wave vector, $k_p(f)$, with frequency. The separation of the input and output CPW, $r$, defines the proportionality constant. But, because the mode dependent, constant phase, $\varphi_{0p}$, is unknown, we can only measure changes in $k_p(f)$. We adjust $\varphi_{0p}$ to bring the corresponding data into coincidence with the model calculation based on Arias and Mills [20,21] for the different lengths and magnetic fields, shown in Fig. 5. It appears that we can only detect the coupling for the lowest order mode for the 10 μm long wire.

¶ The frequency dependent phase extracted from the "noise-like" but reproducible data can be successfully projected onto the Arias and Mills model for spin wave dispersion of several low lying modes. In particular the 5 and 10μm long spin waveguide data can be brought into near coincidence on the dispersion relations in Fig. 5, giving us confidence in this interpretation.

¶ The coupling to various spin waves should be determined by the Fourier components in the exciting field distribution. If the local excitation (and detection) has a spatial scale, $\delta r$, set by the exciting fields near the ends of the CPW and determined by their overlap with the ferromagnetic stripe under the CPW (not by the overall length), we estimate that the most strongly excited spin waves will have Fourier components for spin wavelengths or wave vectors given by $\delta r \sim \lambda/2 = \pi/k \sim 1$ μm. This estimate compares favorably with the range of wave vectors of the modes identified in Fig. 5, approximately 1-5 $10^6$ m$^{-1}$.

¶ By comparing the $|S_{12}|$ amplitude measured for 5 and 10 μm long wires we estimate field attenuation lengths of only $l \sim 3$ μm. Following the model developed by Almeida and Mills [22] we estimate eddy current losses for our 110 nm thick film and find that they are 2-3 orders of magnitude *too small*, to account for this apparent decay length of ~ 3μm. On the other hand ferromagnetic resonance linewidths for the unprocessed films are ~ 0.5 GHz. Combined with the group velocity deduced from Fig. 5 we estimate attenuation lengths by $l \sim |d\omega/dk| \cdot \tau$ of order of ~ 1-2 μm and reasonably close to that deduced from experiment. *It appears that the spin wave decay is caused by spin or magnetization relaxation rather than eddy current damping.*

In summary, we have coupled high frequency coplanar waveguides by magneto-static spin waves in ferromagnetic CoTaZr metal wires; spin wave dispersion appears as a frequency and length dependent phase shift. Attenuation lengths are approximately 3 µm's and much shorter than predicted by eddy current damping. Using the Arias and Mills calculation of spinwave dispersion for infinitely long "wires", we identify up to three different spin wave modes. Magnetic field bias dependence indicates that the spinwave excitation in these wires is sensitive to magnetization alignment and disorder.

These results indicate that spin wave interferometers fabricated with these materials will need to be of the order of or less than microns but magnetostatic spin waves can be "switched" or controlled by local magnetization.

This work is supported by Nano Electronics Research Corporation (NERC) via the Nanoelectronics Research Initiative (NRI), by Intel Corp. and UC Discovery at the Western Institute of Nanoelectronics (WIN) Center.

Figure captions:

Fig. 1  SEM image of a (10×1.8×0.11) μm$^3$ CoTaZr spin wave-guide excited and detected by shorted coplanar waveguides. Inset: Cross section through spin wave guide.

Fig. 2  (Color online) Schematic cross section through the coplanar waveguide above the ferromagnetic metallic wire.

Fig. 3  Frequency dependence of the real part of $S_{12}$ for a 5μm long CoZrTa wire at different longitudinal bias magnetic fields.

Fig. 4  (Color online) a) Frequency dependence of the real and imaginary parts of $S_{12}$, b) polar plot of $S_{12}$ amplitude and phase, and c) Cartesian plot of $S_{12}$ amplitude and phase, measured at $H$=594 Oe on the structure with a (5×1.8×0.11) μm$^3$ CoTaZr wire.

Fig. 5  (Color online) Measured dispersion for 5 μm long (crosses) and 10 μm long (squares) CoTaZr wires with (1.8×0.11) μm$^2$ elliptical profile at 594 Oe. Calculated dispersion (solid lines) for elliptical cross section wires [20, 21]. Left : Bias field 594 Oe. Right: Bias field 164 Oe.

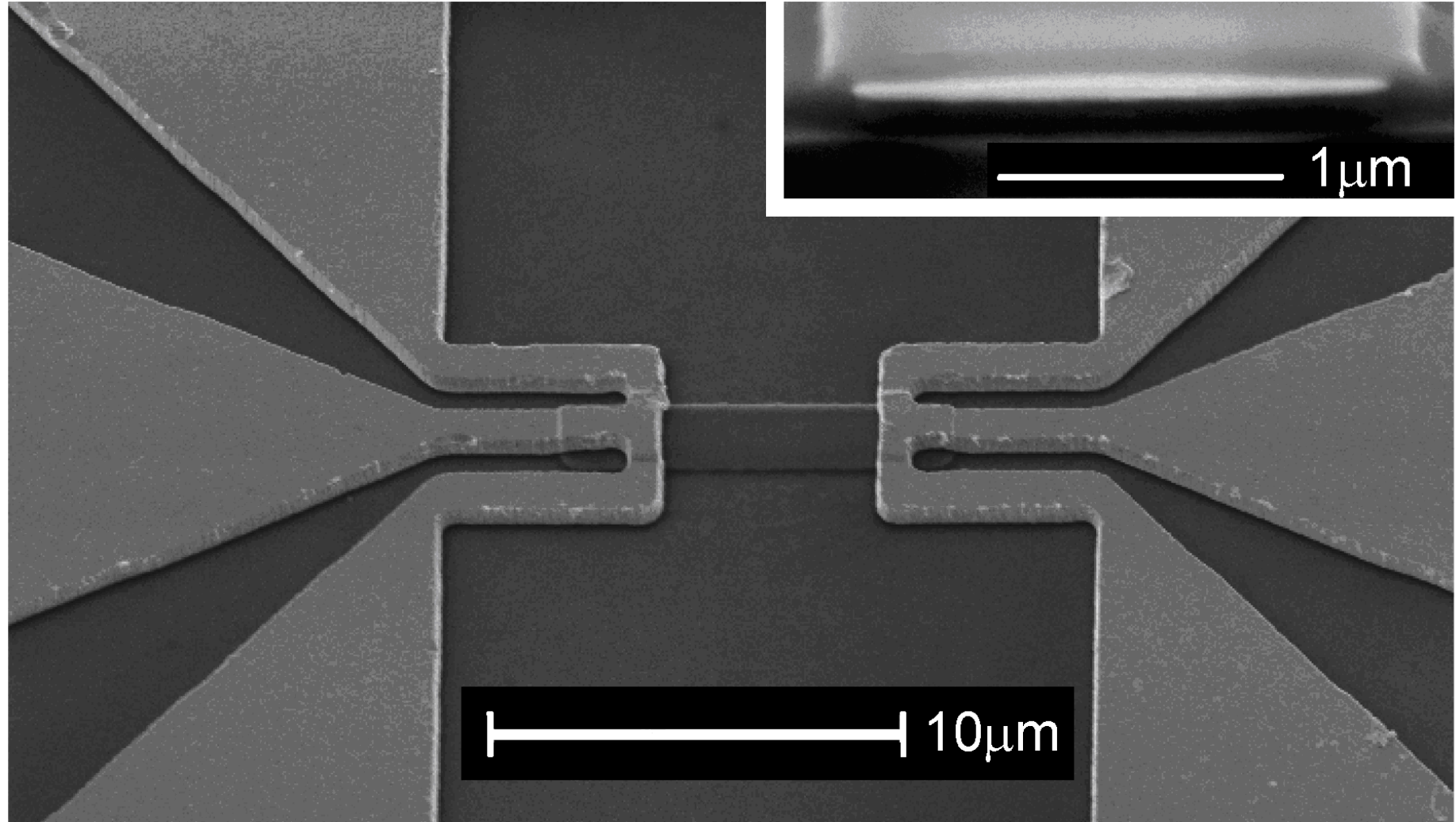

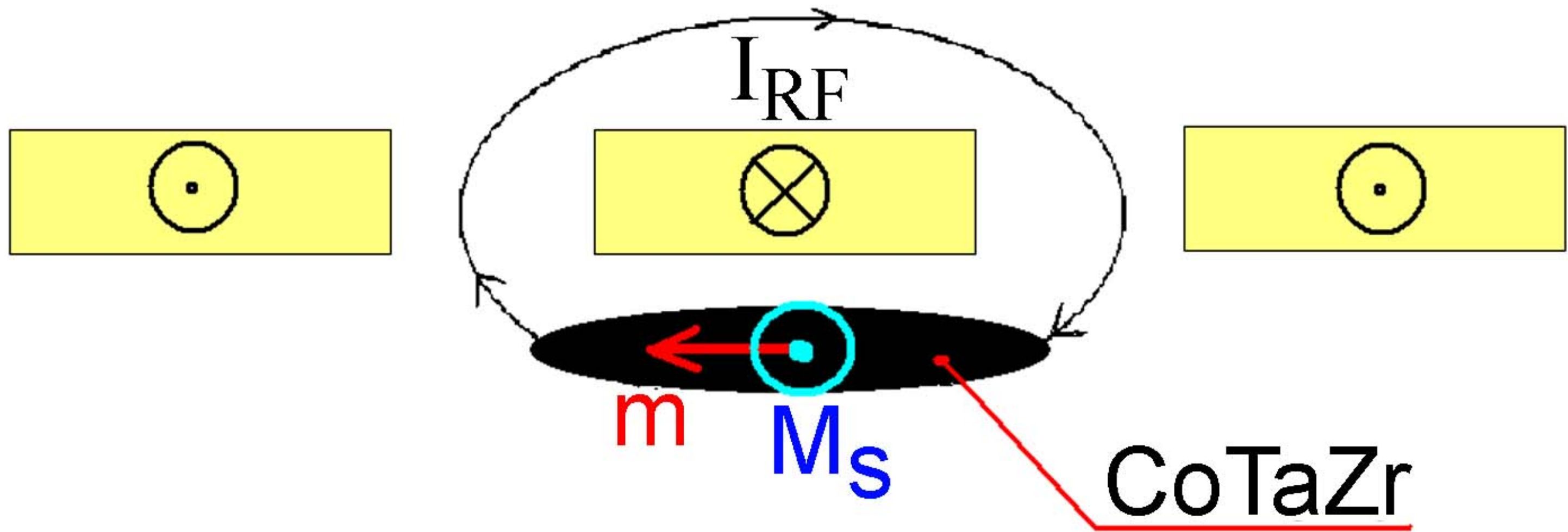

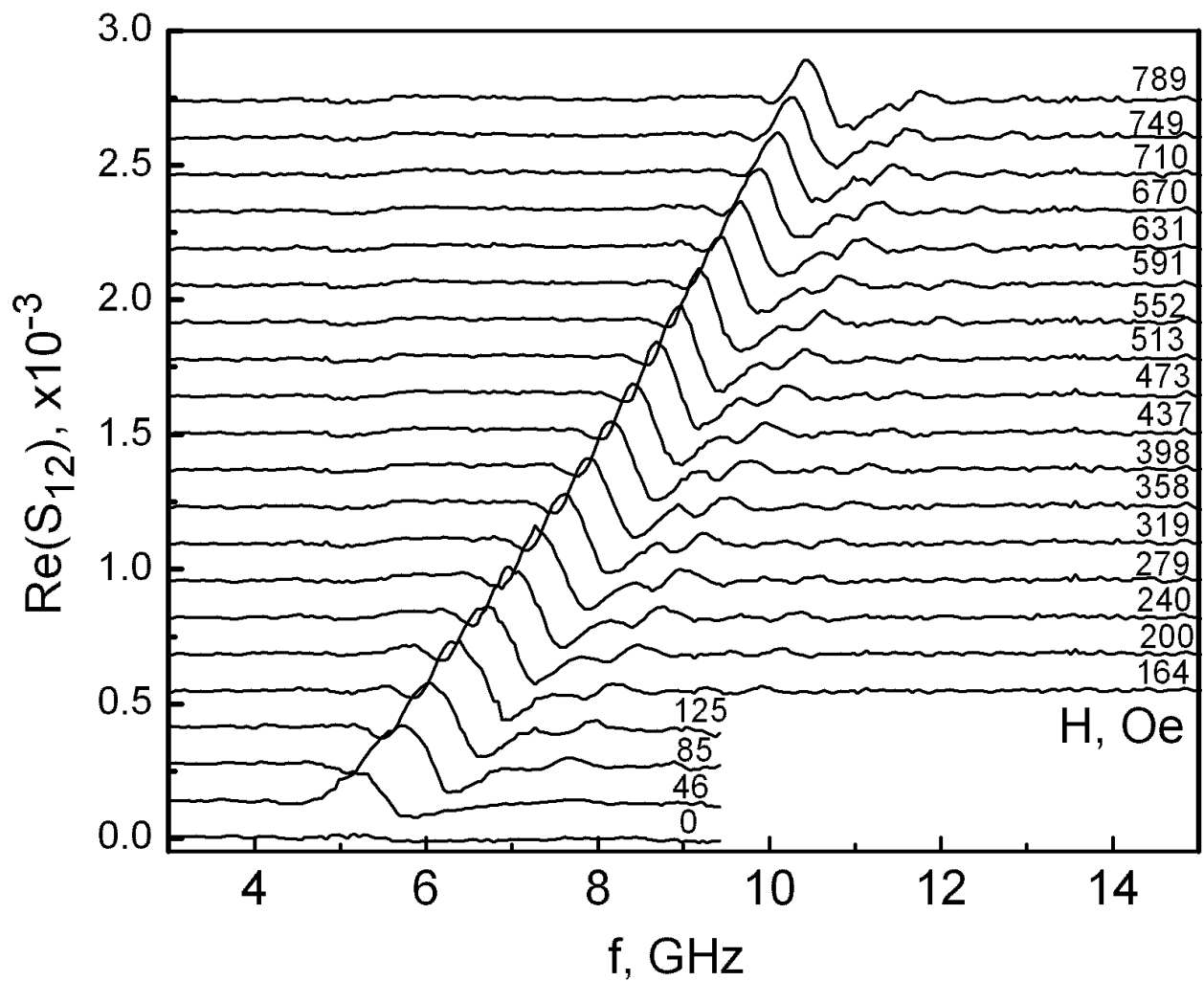

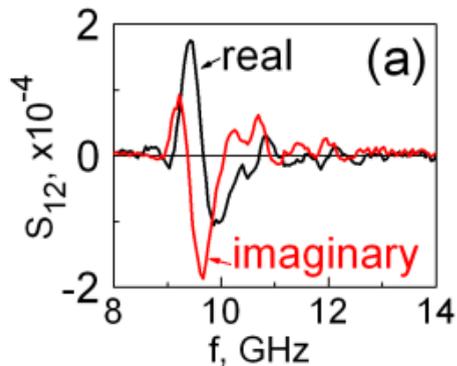
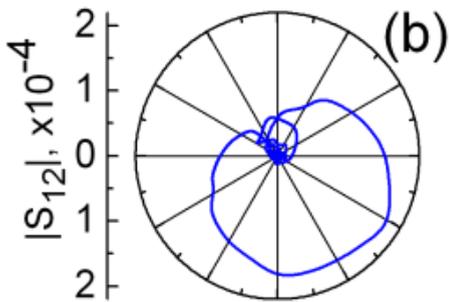
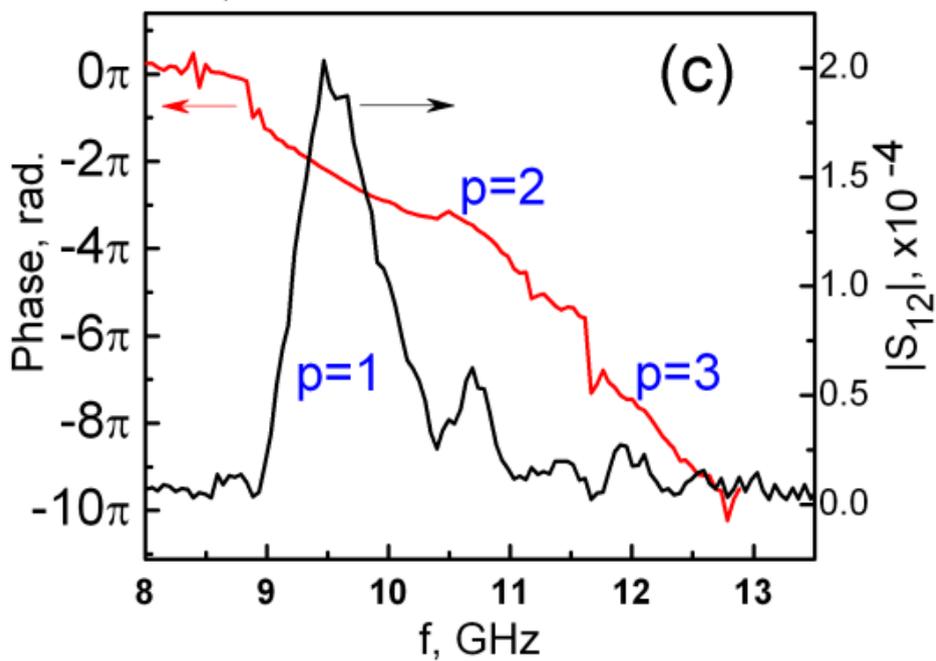

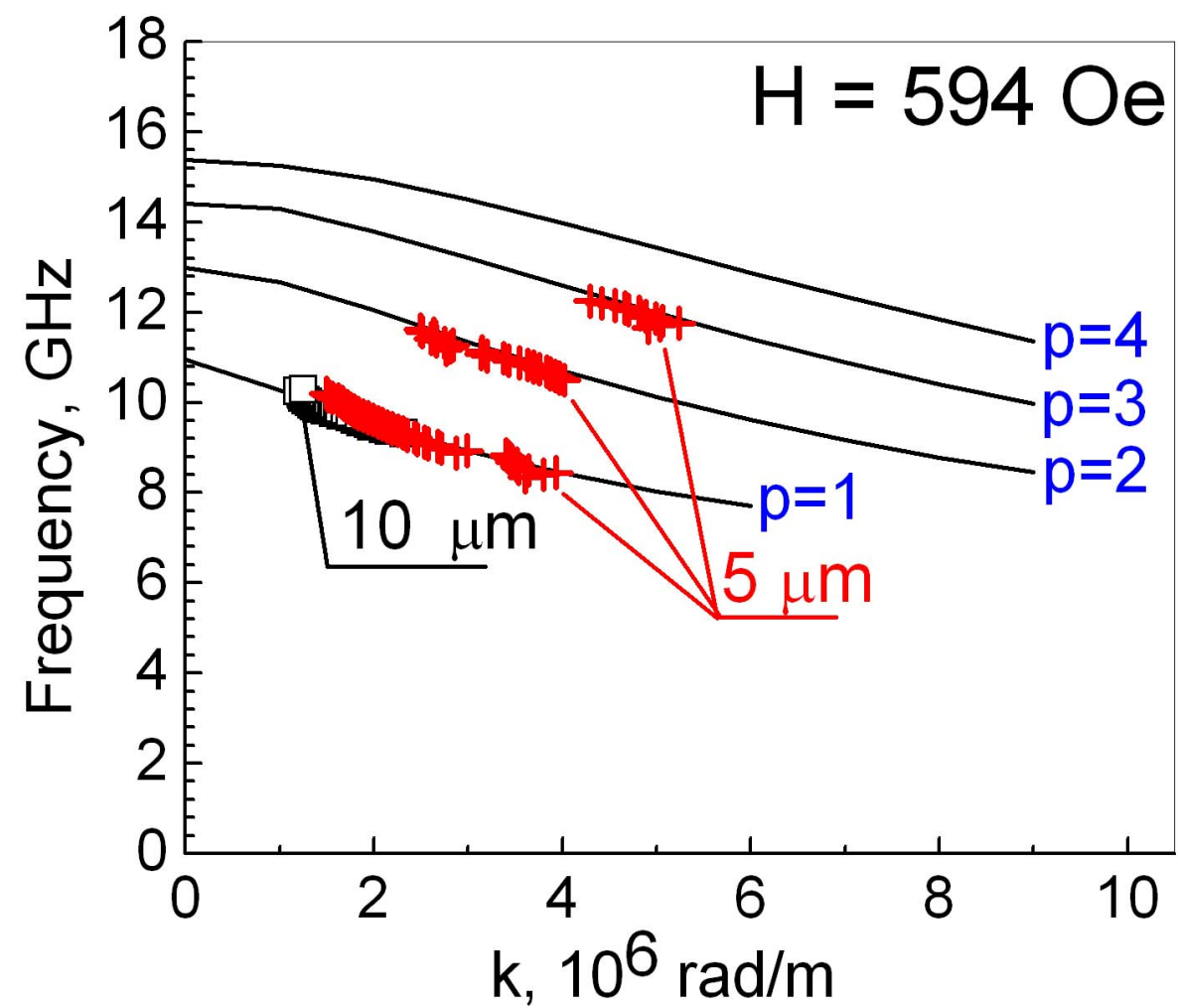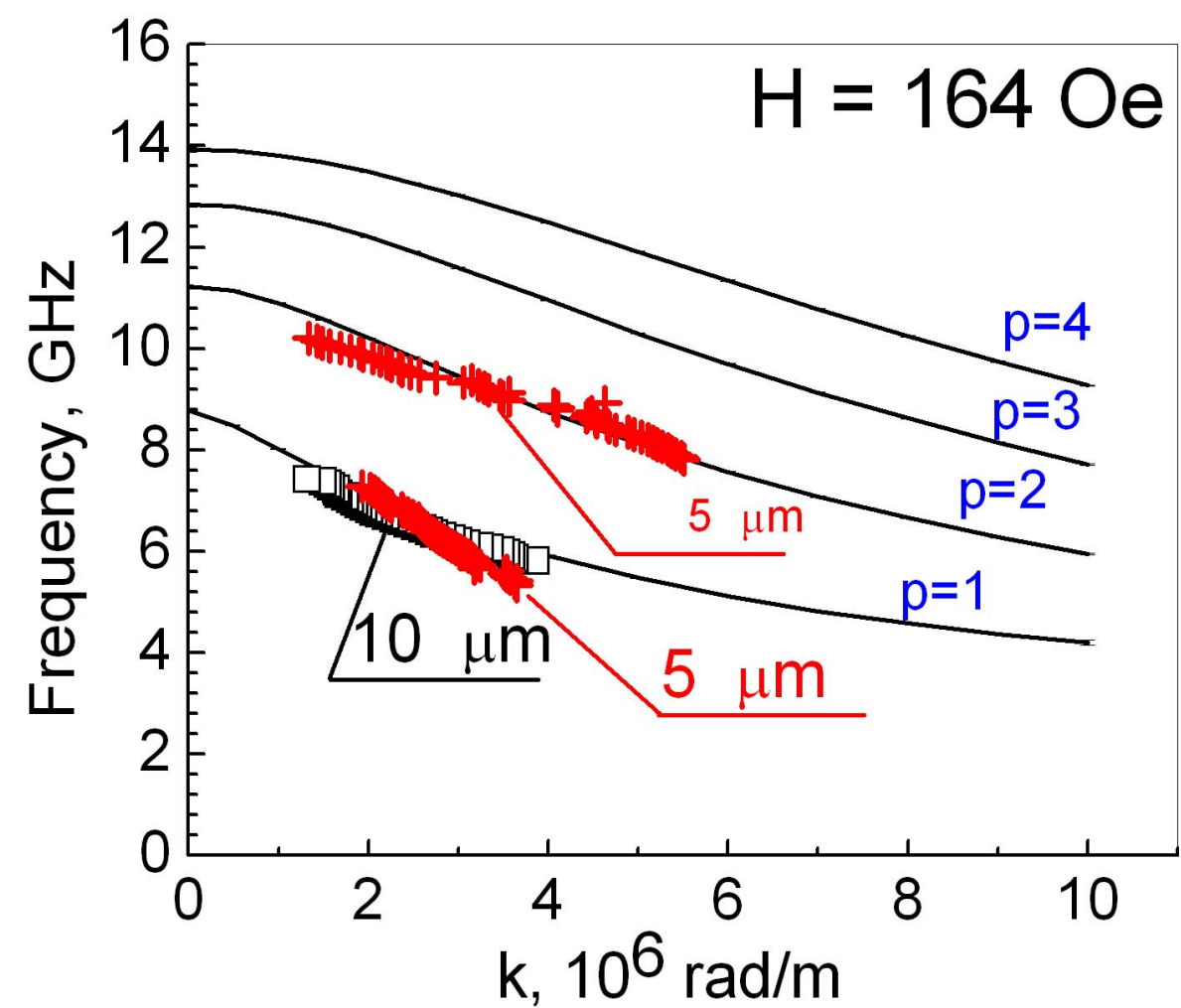